\documentclass[runningheads]{llncs}

\usepackage{amsfonts}

\usepackage{graphicx}
\usepackage[hidelinks]{hyperref}
\usepackage{xcolor}
\usepackage{float}
\usepackage{listings}
\usepackage{multirow}
\usepackage{hhline}

\begin{document}

\title{Duplicate Detection as a Service}

\author{Juliette Opdenplatz\inst{1} \and
Umutcan \c{S}im\c{s}ek\inst{1} \and
Dieter Fensel\inst{1}}

\authorrunning{J. Opdenplatz et al.}

\institute{University of Innsbruck, Technikerstrasse 21a 6020 Innsbruck, Austria
\email{firstname.lastname@sti2.at}}

\maketitle

\begin{abstract}

Completeness of a knowledge graph is an important quality dimension and factor
on how well an application that makes use of it performs.
Completeness can be improved by performing knowledge enrichment.
Duplicate detection aims to find identity links between the instances of
knowledge graphs and is a fundamental subtask of knowledge enrichment.
Current solutions to the problem require expert knowledge of the tool and the
knowledge graph they are applied to.
Users might not have this expert knowledge.
We present our service-based approach to the duplicate detection task that
provides an easy-to-use no-code solution that is still competitive with the
state-of-the-art and has recently been adopted in an industrial context.
The evaluation will be based on several frequently used test scenarios. 


\end{abstract}

\section{Introduction}
\label{section:introduction}

The Knowledge Graph (KG) lifetime contains a cycle of maintenance which is
called KG curation \cite{fensel2020}.
This cycle consists of an assessment, an enrichment, and a cleaning step.
Knowledge enrichment (KE) tasks aim to identify missing statements in KGs.
A special case of KE tasks is called duplicate detection (DD) and is about
finding instances within KGs that describe the same object of discourse.
Solutions to this task have multiple use cases in various fields like linking
medical records \cite{sauleau2005medical} or applications in plagiarism detection
\cite{chowdhury2018plagiarism}.

The fact that DD is such a prevalent task across multiple domains, indicates
that there is a need for easy-to-use DD tools.
Such tools should be operable by non-expert users.
Creating a configuration for a badly documented tool or the manual creation of
training datasets are usually extremely tedious tasks at best.
Simplification of such processes can be achieved by task automation that
abstracts away the parts that require knowledge and time investment from
users.

Our approach, Duplicate Detection as a Service (DDaaS) aims to provide a
solution to the DD task that does not require any expert knowledge of its users.
It utilizes the schema of a KG automate all the behind-the-scene processes and
almost all processes that are usually left to the user.
An active learning approach is applied to train DDaaS for a KG.
The training process requires user to make yes/no decisions about proposed
duplicates.
This is the only action that is really required to be done by the user.
A demonstration is provided\footnote{\url{https://ddaas.sti2.at}}.

The remainder of this paper is structured as follows: Section
\ref{section:approach} discusses the approach DDaaS implements.
This implementation is described in Section \ref{section:implementation}.
The state-of-the art to this paper is then presented in more detail in Section
\ref{section:related-work}.
In Section \ref{section:conclusion-and-future-work}, we conclude the paper and
explain our plans for the future of DDaaS.

\section{Approach}
\label{section:approach}

\begin{figure}[ht]
    \centering
    \includegraphics[width=\textwidth]{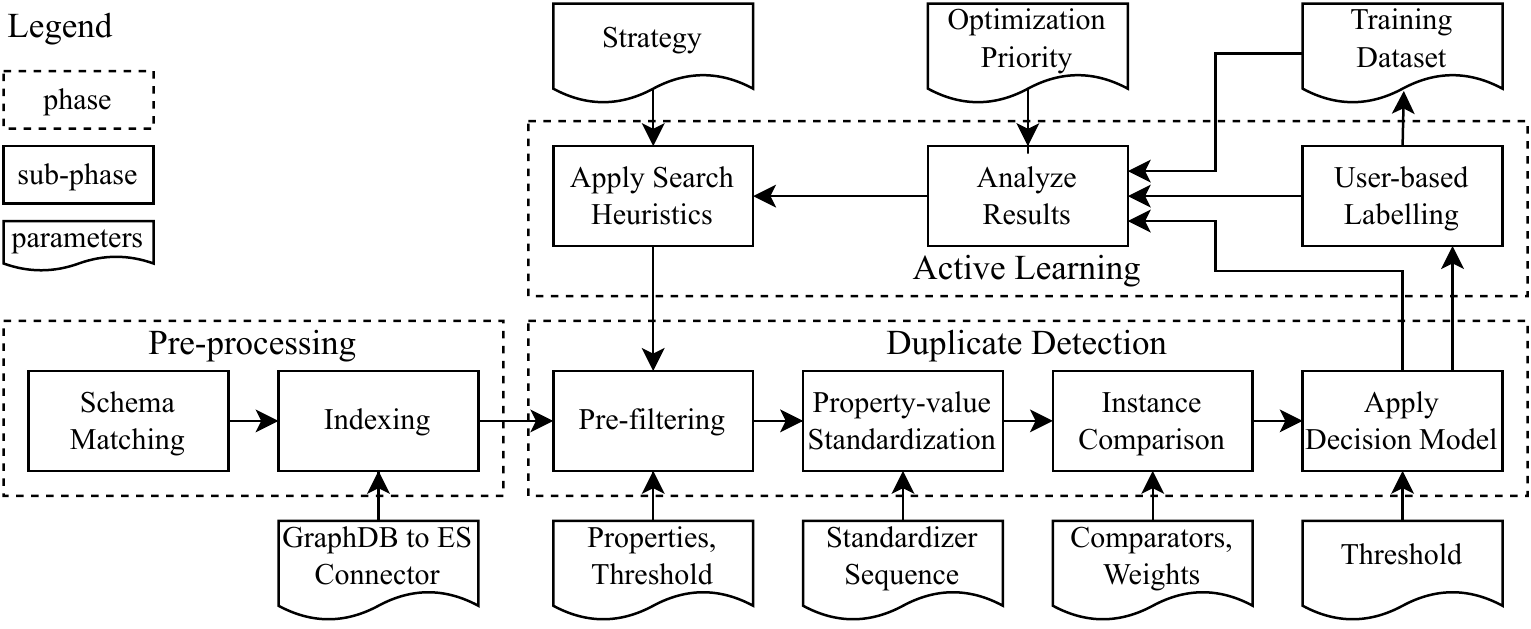}
    \caption{Overview of the DD process, active learning, and all necessary configurations.}
    \label{figure:overview-dd-process}
\end{figure}

In this Section, we give an overview on the approach that DDaaS implements.
We divide the approach in three different phases:
(1) A pre-processing phase that is a prerequisite to performing DD.
This is described in Section \ref{subsection:approach-pre-processing}.
(2) The DD phase which is described in Section
\ref{subsection:approach-duplicate-detection}.
(3) The active learning phase which is built on-top of the output of the DD
phase.
A detailed overview, is given in Section
\ref{subsection:approach-active-learning}.
These phases consist of sub-phases as depicted in Figure
\ref{figure:overview-dd-process}.

\noindent
As displayed, there are different parameters to each sub-phase.
In the following we give an explanation of these parameters for each sub-phase
in the respective section.
This explanation also consists of a description of an estimate of possible
values, which we call the solution space to give an overview of the complexity.

\subsection{Pre-processing}
\label{subsection:approach-pre-processing}

The pre-processing phase is composed of two different sub-phases (1) schema
matching which aims to make different KGs comparable and (2) indexing which
enables fast access to the relevant instances of the KGs.
These sub-phases are further described in this section.

\subsubsection{Schema Matching}
Schema matching is the task of aligning KGs by mapping them to a common schema.
It is worth noting, that only a mapping between types and the properties of
interest is required.
This common schema allows for DDaaS to determine what instances and what
properties are comparable.
As DDaaS does not provide any schema matching capabilities, there are no
parameters for this sub-phase.

\subsubsection{Indexing}
In this sub-phase, instances and property-values of interest are stored in an
index.
An index stores instances of a single type of the KG in a special data
structure that allows for quick access and search of instances.
DDaaS stores property-values up to a specified depth following an instances'
property paths.
Structures like inverted indices are used for text indexing and BKD-trees
\cite{Sharma2020} are used for indexing numerical values.
The capability to search for instances quickly is crucial for run-time reasons.
The parameters of this pre-processing sub-phase are dependent on technologies
used and therefore, we refer to Section \ref{subsection:implementation-indexing}
where we explain how the indexing is implemented.

\subsection{Duplicate Detection}
\label{subsection:approach-duplicate-detection}

The DD phase is composed of four different sub-phases which loosely follow a
commonly used process model that was previously described by Baxter et al.
\cite{baxter2003a-comparison-of-fast-blocking-methods-for-record-linkage}.
We have four sub-phases: (1) Pre-filtering which builds on top of the indexing
phase, (2) property-value standardization which aims to make values more
comparable, (3) instance comparison where instances are compared property-value
by property-value, and (4) decision model application where we decide whether an
instance comparison results in a proposed duplicate.

\subsubsection{Pre-filtering}
This sub-process intends to find candidate duplicates for each instance of a KG
in another or the same KG.
It does so by iterating over all instances of one index (\texttt{source}) and
pulling candidate duplicates from another specified index (\texttt{target})
using a set of properties/property paths from the common schema and a percentage
value that declares the amount of terms that have to match between two
instances' property-values.
These requirements form the parameters to the pre-filtering phase:
\begin{enumerate}
    \item\textbf{Property Selection:}
    A solution is a non-empty subset of properties from the common schema.
    The solution space is the set of all these subsets.
    \item\textbf{Threshold:}
    A solution is a number in the range $[0,100] \in \mathbb{N}$.
    This range is the solution space.
\end{enumerate}
\noindent
The result of this phase is used for a more detailed comparison.

\subsubsection{Property-value Standardization}
Standardization of property-values of a KG transforms these values into a
uniform format.
This makes their comparison more consistent.
For example, string values might be more comparable when they are consistently
represented in lowercase format or numbers might be more comparable when they
use the same amount of decimals.
We differentiate between multi-value and single-value standardization.
Multi-value properties are standardized on two different levels:
(1) The list level and (2) the element level.
Elements are standardized before the list level standardization is applied in
order to make the list level standardization more efficient. 
The standardization phase is configured by listing all the standardization
functions that shall be applied to each property. 
This leads to the following parameters:
\begin{enumerate}
    \item\textbf{Property Selection:}
    A solution is a non-empty subset of properties in the common schema.
    The solution space is the set of all these subsets.
    This makes the problem equivalent to Pre-filtering Property Selection
    which indicates re-usability of methods.
    \item\textbf{Standardization Functions:}
    A solution is a mapping of the subset of properties from the common
    schema to a sequence of standardization functions.
    There are two problems with this:
    (1) Each sequence is potentially unlimited in size as we can
    theoretically apply an unlimited amount of standardization functions.
    (2) Some of these functions are parameterized with an unclear or
    infinite range.
    As a result we can not provide a limited solution space.
\end{enumerate}
\noindent
After the standardization phase, the standardized instances are passed to the
instance comparison phase. 

\subsubsection{Instance Comparison}
\label{subsubsection:instance-comparison}
The instance similarity is computed by taking the weighted average similarity of
two instances' property-values.
For this to work we need to define measures to determine the similarity of two
property-values.
This can be a tricky matter as have to differentiate multiple cases depending
on the range(s) of a property:
\begin{enumerate}
    \item \textbf{Literal to literal:} Property-value similarity measures take
    as input two property-values $a$ and $b$, their output has to be in the
    range of $[0, 1]$ where values close to $0$ indicate low similarity and
    values close to $1$ a higher similarity.
    Examples for such a measure include the good old levenshtein similarity for
    string values, the ratio for number values, or a measure that checks whether
    two boolean values are the same.
    \item \textbf{Instance to literal:} Property-values that can be represented
    by literals and instances usually provide the same information in both
    forms.
    For the instance form, the information is likely distributed across multiple
    properties.
    To make these comparable, we indexed whole property paths for an instance to
    make them comparable with the functions for literal to literal comparison.
    \item \textbf{Instance to instance:} We have multiple options for this type
    of comparison:
    (1) Comparison of the URIs that are used to identify the instances.
    (2) Direct value comparison of the stored property paths and treating them
    just like direct property-values.
    (3) Individual comparison of their properties, and aggregation of their
    similarities.
    Options (1) and (2) make this type of comparison a literal to literal
    comparison while option (3) is the most sophisticated way of comparing
    instances but also requires the most effort in run-time, configuring, and
    understanding.
\end{enumerate}

\noindent
In addition to the distinction of literal values and instance identifiers,
a differentiation has to be made between single- and multi-value comparisons.
For those, we first compute similarities between the elements of the lists and
aggregate the resulting similarity scores.
This aggregation can consist of simply taking the highest, average, or lowest
similarity for all possible comparisons between two lists of property-values.
Since we give weights to each property and property path, we have the following
parameters:
\begin{enumerate}
    \item\textbf{Property Selection:}
    A solution is a non-empty subset of properties in the common schema.
    The solution space is the set of all these subsets.
    \item\textbf{Comparison Functions:}
    A solution is a mapping of the property selection.
    The solution space for comparison measures is every possible mapping.
    \item\textbf{Weights:}
    A solution is a mapping of the property selection to $[0,1] \in \mathbb{R}$
    limited to two decimals.
    The solution space is every possible mapping.
\end{enumerate}
\noindent
As a result of the instance comparison phase, each pair of instance and
candidate duplicate is assigned a similarity value.

\subsubsection{Decision Model}
During the decision phase, the instance pairs are filtered by applying a
similarity threshold $t \in [0, 1]$.
This \textbf{decision threshold} is the only parameter to the decision model.
A solution is a number in the range $[0,1] \in \mathbb{R}$ limited to two
decimals.
This range is the solution space.
Every instance pair above this threshold is a proposed duplicate.
This finalizes the DD phase.

\subsection{Active Learning}
\label{subsection:approach-active-learning}

The active learning phase utilizes the resulting proposed duplicates of the DD
phase.
It contains three sub-phases which are described in this section:
(1) user-based labelling,
(2) result analysis,
and (3) search heuristics application.

\subsubsection{User-based Labelling}
\label{subsubsection:user-based-labelling}

A learning process requires a training dataset.
Such a training dataset consists of instance pairs and a label that declares a
pair as duplicate (\texttt{true}) or non-duplicate (\texttt{false}).
Use-cases usually do not have such a dataset in advance.
The user is required to label the partial result of a DD phase.
This phase does not have any parameters.

\subsubsection{Result Analysis}
\label{subsubsection:result-analysis}

To determine whether a configuration is better than another, we provide three
well-known measures to optimize for: (1) Recall, (2) Precision, and (3)
F1-measure.
These measures are used as a configurable parameter for the learning process.
We specify a primary parameter, that declares what measure shall be mainly
optimized for.
If this value is equal for two setups, we optimize for the secondary parameter.
In order to perform such a result analysis, apart from the quality measures, a
training dataset is required.
The result consists of the quality measures described above.

\subsubsection{Search Heuristics}
\label{subsubsection:search-heuristics}

Now that the quality of a DD result can be quantified, different setups can be
compared.
It also enables the use of search heuristics to find better setups by
iteratively changing the value of DD parameters.
Such heuristics include Forward Selection, Backward Elimination, Hill-Climbing,
and Genetic Algorithms.

For each DD parameter, there is also a brute force option which is viable if the
solution space for that DD parameter is small enough.
Forward selection for property selections works by measuring the results for
each property in isolation, the properties are sorted by quality of result, and
incrementally added to the configuration until the configuration starts to get
worse.
For the property comparison configuration this means the outvoted properties'
weights are set to $0$.
Backward elimination works the other way around, each property is evaluated in
isolation and the worst properties are incrementally ruled out.
Hill-climbing algorithms work by step-wise increase/decrease (depending on where
the result gets better) of the numerical value of the percentage, threshold, or
weight parameters.
Genetic algorithms are used to randomly change comparison and standardization
functions for a property.
The order and amount of times each of these algorithms is applied, is
configurable.
This is called the search strategy which is the parameter to the search
heuristic application sub-phase.

Since the strategy is a sequence of search heuristic applications, each of
which includes running the DD phase once, the execution of a strategy is not
interrupted by user-based labelling.
Only after the strategy is completed, the user is prompted to label results
again.

\section{Implementation and Running Example}
\label{section:implementation}

\begin{figure}[ht]
    \centering
    \includegraphics[width=\textwidth]{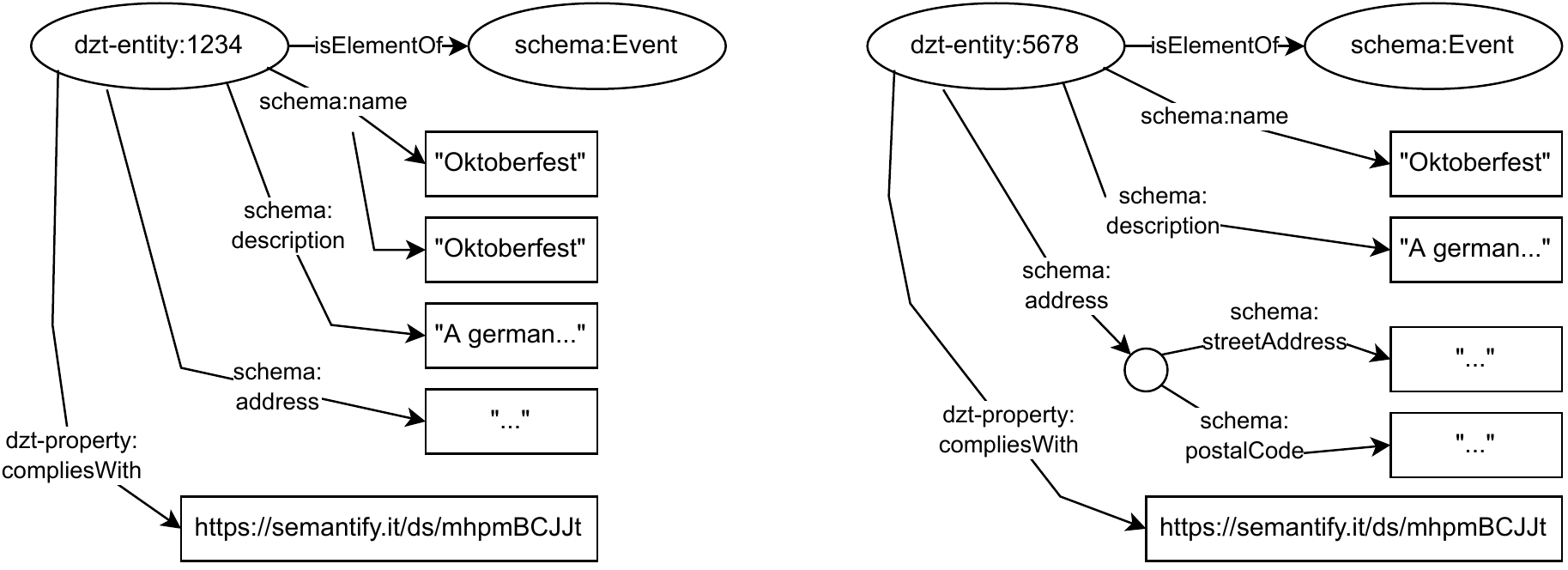}
    \caption{Running example for two instances of the \texttt{schema:Event} type.}
    \label{figure:running-example}
\end{figure}

Since the goal of DDaaS is to provide a user-friendly and accessible approach to 
DD, lots of automation has to be performed.
To showcase the implementation and automation of phases, we provide a running
example as per Figure \ref{figure:running-example}.

\noindent
To achieve automation, means to set parameters to the approach as automated as
possible.
To do so, DDaaS makes use of the schema that is provided by a KG.
In this Section we discuss the implementation details to the presented approach
with a special focus on how most phases are automated.

In the following, we compare two instances of type \texttt{schema:Event}.
Both are based on the same domain specification (see \texttt{compliesWith}
property).
Additionally, they both contain values for a \texttt{name}, \texttt{description},
and \texttt{address} which is differently structured for each instance.
The left entity also has multiple values for the \texttt{name} property whereas
the one on the right only has one value for the same property.
These graphs are stored in a 
GraphDB\footnote{\url{https://www.ontotext.com/products/graphdb/}} triple store.

A more detailed description of each phase's implementation is provided in this
section.
Section \ref{subsection:types} highlights the importance of structural
KG information for automation in DDaaS.
In Sections \ref{subsection:implementation-indexing} and
\ref{subsection:implementation-pre-filtering} it is described how the
indexing is automated and how candidate pairs are generated by utilizing these
indices.
We give an overview of the DD phase implementation in Section
\ref{subsection:implementation-dd}.

\subsection{Schema Information Extraction}
\label{subsection:types}

Large parts of the automation of processes can be attributed to the information
that the schema of a KG provides.
DDaaS currently implements extraction of schema information for process
automation in two different ways: (1) From SHACL-based domain specifications
\cite{fensel2020,csimcsek2017domain} which usually comes with the graph itself
in form of a URI to the SHACL-formatted resource.
The domain-specifications are patterns of extended subsets of schema.org to
facilitate knowledge creation.
Local properties and their constraints are defined on schema.org types and its
extensions \cite{simsek2021knowledge}.
(2) From emergent schemata \cite{hogan2021knowledge} where the schema
information is inferred directly from the statements of the KG by using SPARQL
queries.
Either way the following information is gathered:
(1) Types and respective property paths,
(2) Property cardinalities, and
(3) Literal Datatypes.

We exploit knowledge about the XML schema datatypes
(XSD)\footnote{\url{https://www.w3.org/TR/xmlschema-2/\#built-in-datatypes}}
of a KG.
XSDs are usually the go-to-datatypes for literals in KGs.
For instance, lots of these XSDs are categorized as \texttt{dd:String}, like
\texttt{xsd:string} or \texttt{xsd:anyURI}.
A likewise mapping is done for number and boolean types.

This list can be extended in case a KG uses other labels.
There is also a loss of granularity by mapping all the time-related XSD merely
to the \texttt{dd:String} type.
The gathered information, independent of whether it came from a DS or the
emergent schema, is stored in an internal minimal domain specification which is
used to generate all kinds default configurations.

\subsection{Indexing}
\label{subsection:implementation-indexing}

During the indexing phase, instances stored in the triple store are indexed in
an Elasticsearch\footnote{\url{https://www.elastic.co/elastic-stack/}} cluster.
The instances are stored in a flat \texttt{JSON}-like document format.
Conveniently, in its enterprise edition, GraphDB comes with a dedicated
connector functionality that automates the indexing process and keeps the
created index up-to-date.
Such a GraphDB to Elasticsearch Connector
(GEC)\footnote{\url{https://graphdb.ontotext.com/documentation/enterprise/elasticsearch-graphdb-connector.html}}
can be configured using the Workbench or a SPARQL query.
We create one index per type in the KG which means that there is a GEC generated
for each type.

\subsubsection{Running Example.}
In case of our running example, we create a GEC for the internal minimal domain
specification that was obtained in SHACL format by following the URI value of
the \texttt{compliesWith} property.
Any property that has another non-literal type as range, maps the property
paths as fields to the index as follows:
\texttt{schema:address} simply becomes \texttt{address} and the property path
\texttt{schema:address/schema:postalCode} is mapped to a field called
\texttt{address.postalCode}.
Additionally, one can specify whether a property is a multi-value property to
the GEC, and therefore, enforce the values of this field to be stored into an
array.
This information is also taken from the domain specification.
In our running example this is the case for the \texttt{name} property.
This process is repeated for the specified property path depth.
The default depth is set to $1$ which results in an index where only a second
layer is added to the index.
A default depth is applied since we can not rule out that the graph does not
contain cycles, and we end up with an infinitely large configuration.
The resulting generated GEC is uploaded via the GraphDB's SPARQL endpoint.

After the process is finished, we end up with indexed instances, which to the
outside are represented by a flat \texttt{JSON} document that lists instances'
property paths and their values.

\subsection{Pre-filtering}
\label{subsection:implementation-pre-filtering}

The pre-filtering approach is based on so called \texttt{more\_like\_this}
(MLT)\footnote{\url{https://www.elastic.co/guide/en/elasticsearch/reference/current/query-dsl-mlt-query.html}}
queries which finds instances that look similar to a sample instance with
respect to the following parameters: (1) property selection and (2) threshold.
For (1), we simply utilize every available indexed property (path).
The \texttt{source} index provides the samples which are requested via a query
that returns all stores instances, these instances are used as the sample for
the MLT query.

\subsubsection{Running Example.}
The resulting default property selection for the running example looks like:
\texttt{[address, address.postalCode, ...]} and includes every possible
property (path).
We also specify an arbitrary default value of $40$\% for (2).
Iterating over the \texttt{source} index's instances and taking them as
samples for the MLT query the resulting query is then sent to the respective
Elasticsearch endpoint which returns a list of candidate duplicate instances
from the \texttt{target} index.
So, the pre-filtering step would recognize for \texttt{dzt-entity:1234} that a
similar instance called \texttt{dzt-entity:5678} exists and return that.

\subsection{Instance Standardization, Comparison, and Decision Model}
\label{subsection:implementation-dd}

The parameters for instance standardization, comparison, and decision model form
a DD configuration.
Standardization and comparison functions are applicable only to values of
certain literal datatypes.
To generate defaults for a property, we apply default functions depending on
their literal datatype.

\subsubsection{Running Example.}
We apply the \texttt{lowercase} standardizer on each value for the \texttt{name}
property and compare each value of one instance to each value of the other
instance and choose the highest similarity as the similarity score for this
property.
The \texttt{description} property results in a simple string comparison.
A little more complicated is the comparison of both \texttt{address} values as
they differ in structure for the running example.
In such cases, we serialize the values of the property paths into a single
string as needed during the DD process and compare those values. 
Whether a value is an instance or not is determined by checking the property
paths' fields.
If they are empty, the property-value is a literal, if they are not empty the
property value is an instance, and they are serialized.
The result is a simple string comparison as if both were literals.
If both property-values are instances, we simply compare each property path
individually.
For the weights we simply assign a value of $1$ to each property since it is
hard to determine their importance automatically.
The threshold value is initially arbitrarily set to $0.75$.

Using this configuration, we compare the instances \texttt{dzt-entity:1234} and
\texttt{dzt-entity:5678} in more detail.
We compare the multi-value property \texttt{name} which for both instances
contains only one value due to the application of the \texttt{setify}
standardizer.
By default we select the most similar pair of property-values, which is now
trivial.
The elements are compared using \texttt{levenshtein} which is the default
comparator for string typed values.
For the \texttt{description} property, we have a simple string comparison.
In case of the \texttt{address} property of \texttt{dzt-entity:1234} we see that
the sub-properties are populated, whereas for the other instance they are not.
Thus, we serialize the sub-properties of the \texttt{source} entities into a
single string which is then compared to the other address as a simple string
comparison.
The \texttt{compliesWith} property is ignored.
The result of this comparison is a list of similarities.
Due to the default weights being $1$ for each property, we take simply the
average similarity and check with the similarity threshold whether the two
instances can be considered duplicates.

\subsection{Active Learning}
\label{subsection:active-learning}

The training dataset for an index pair is stored in a
MongoDB\footnote{\url{https://www.mongodb.com/}} database and consists on the
pair of URIs of two instances and a boolean property that indicates whether the
instances are duplicates.
When there was not already any training data provided for the index pair, we
run the duplicate detection process on the automatically generated default
configurations as per Sections \ref{subsection:implementation-pre-filtering} and
\ref{subsection:implementation-dd} and set the decision model threshold very
low.
The search strategy is applied as explained in Section
\ref{subsection:approach-active-learning}, where we use the newly created
training data to analyze the results.
During this search phase, the user-based labelling is skipped until the whole
strategy was applied.
After the whole strategy was applied, the user is prompted to label new pairs,
which were not yet labelled.
Whether a pair was already labelled is simply looked up in the database.

\subsubsection{Running Example.}
If our running example was part of a completely new DD process, there might not
have been a training dataset, therefore the result of the comparison of our two
instances would have been prompted to the user for validation sooner or later
(depending on other examples in the same KG).

Depending on the outcome of other possible comparisons, the result analysis
might result in a more or less bad result.
The search heuristic strategy would be applied and iteratively mutate one
parameter at a time for the DD phase until the whole strategy was applied.
Then the cycle of labelling, analysis, and search would continue until the
result is satisfying for the use-case.

\section{Related Work}
\label{section:related-work}

We discuss and compare the presented approach to the state-of-the-art in the DD
field.
This section is limited to approaches that are specifically designed for KGs.
We compare following tools:
\begin{enumerate}
\item 
\textbf{Duke}\footnote{\url{https://github.com/larsga/Duke}} \cite{garshol2013}
is a tool that is not primarily aimed to DD for KGs, therefore it supports
JDBC databases and file formats like CSV and JSON.
The approach is very similar to ours, and also includes active learning
capabilities.
However, it requires to do the configuration manually which includes the
connection to the database, the individual parameters for each property.
These require detailed knowledge on about every aspect of the data.
It also lacks the capability to properly compare multi-value properties.
Even though it is capable to compare property paths, the functionality
does not work in a flexible manner.
\item 
\textbf{SILK}\footnote{\url{https://github.com/silk-framework/silk}}
\cite{volz2009}
does not only focus on detecting identity links but also other types of RDF
links and works in use-cases where different and inconsistent vocabularies
are being used.
It comes with a flexible language which is utilized to configure DD processes.
Users have to specify multiple different things that refer to implementation
details like SPARQL endpoints, whether it should cache instances, output files,
and page sizes for example.
Users have to specify property paths themselves.
Even though it is very flexible, it does not seem to differentiate between
property-values for a property that are of different types.
The comparison process and decision models work very similarly to our
approach and even offer more flexibility in some areas such as the pre-filtering
options.
\item
\textbf{LIMES}\footnote{\url{https://github.com/dice-group/LIMES}}
\cite{ngomo2011}
is a link discovery tool with a primary focus on large-scale and run-time
requirements by implementing a novel approach to the pre-filtering step.
Other than that its approach for the most part follows the common workflow
described in Section \ref{section:approach}.
It also provides active learning components but does not abstract away its
functionalities such that knowledge about the internals are still necessary
even though their demo
version\footnote{\url{https://limes.demos.dice-research.org/}} is quite
impressive.

\item
\textbf{EAGER}\footnote{\url{https://github.com/jonathanschuchart/eager}}
\cite{obraczka2021embedding}
complements the usual approach described in Section \ref{section:approach} by
graph embeddings \cite{dai2020,wang2017}.
The decision model therefore relies on embedding vectors and attribute
comparisons as input for classification.





\end{enumerate}

\noindent
We summarize our findings in Table \ref{table:dd-tool-comparison} where we list
our main requirements for DDaaS and for each tool evaluate whether it fulfills
each requirement.

\begin{table}[H]
    \centering
    \caption{Comparison of the state-of-the-art in DD.}
    \label{table:dd-tool-comparison}
    \begin{tabular}{|p{0.125\textwidth}||p{0.15\textwidth}|p{0.11\textwidth}|p{0.11\textwidth}|p{0.105\textwidth}|p{0.1\textwidth}|p{0.12\textwidth}|p{0.1\textwidth}|}
        \hline
                          &                     & \multicolumn{3}{c|}{\textbf{Automation Factors}}                        & \multicolumn{3}{c|}{\textbf{Property Comparison}} \\
        \hline
        \textbf{Approach} & \textbf{supports}   & \textbf{learning} & \textbf{active learning} & \textbf{utilizes schema} & \textbf{multi-value} & \textbf{property paths} & \textbf{range-specific}           \\
        \hline                          
        \hline                          
        DDaaS             & RDF                 & yes               & yes                      & yes                      & yes                  & yes             & partially                         \\ 
        \hline                                                                                                                                      
        Duke              & RDF, JDBC, CSV, JSON& yes               & yes                      & no                       & no                   & no              & no                                \\ 
        \hline                                                                                                                                      
        Silk              & RDF                 & no                & no                       & no                       & yes                  & yes             & no                                \\
        \hline                                                                                                                                                     
        LIMES             & RDF                 & yes               & yes                      & no                       & no                   & yes             & no                                \\
        \hline                                                                                                                                                                                         
        EAGER             & RDF                 & yes               & no                       & no                       & no                   & no              & no                                \\
        \hline
    \end{tabular}
\end{table}

\noindent
According to our research, there was no other approach that fulfill all
requirements, we defined for DDaaS.
Most approaches disregard the existence of multi-value properties and focus
purely on single-value properties or merely provide an extremely simplistic
method of aggregation.
Another commonly disregarded point is the existence of different ranges for
a given property.
All the investigated tools will compare literals with URIs when applied
on the wrong KG\footnote{E.g., see \url{https://schema.org/address}}.
For our approach we wanted to provide a similar flexibility that KGs provide to
us as this is the only way that two KGs' instances can be compared properly.

\section{Conclusion \& Future Work}
\label{section:conclusion-and-future-work}

With this paper, we presented a no-code solution to the DD
task that helps non-coders to maintain KGs that power their applications.
This is of great value for any project that aims to keep their KGs up-to-date
without having to get into the technical details of the task.
Hence, we named our solution Duplicate Detection as a Service (DDaaS).
We give a final conclusion on DDaaS in Section \ref{subsection:conclusion} and
an outlook on future plans in Section \ref{subsection:future-work}.

\subsection{Discussion}
\label{subsection:conclusion}

The only necessary configuration DDaaS is not capable of automating is the
necessary schema matching.
More precisely the property and type mapping.
Therefore, DDaaS provides a fully automated default functionality for any KGs
that share a schema.

\subsection{Future Work}
\label{subsection:future-work}

We presented an approach that is usable by non-experts, next steps will focus
on enhancing the DD processes efficiency without increasing the expert-knowledge
that is needed to operate DDaaS.
A first improvement to be made is the setting of default weights for
property-value comparison.
Since we store configurations per index pair and the respective schema, DDaaS
can potentially make use of its experience on new KGs.
One big improvement to be made is the schema matching which still has to be done
completely manually.
Active learning with the help of a well-designed underlying logic could help
create those mappings at least semi-automatically. 
The greater vision for DDaaS is to integrate it into a more general knowledge
enrichment framework that is capable of detecting links of any kind.

\section*{Acknowledgements}

The initial work for this project was funded by the
MindLab\footnote{\url{https://mindlab.ai/}} project.

\bibliographystyle{splncs04}
\bibliography{paper}
\end{document}